\documentclass[ journal=jacsat,
                manuscript=article,
                ]{achemso}
\usepackage{graphicx} % Required for inserting images
\usepackage{siunitx}
\usepackage{nicefrac}

\newcommand{\didv}{$\mathrm{d}I/\mathrm{d}V$}

\title{Vibrational excitations in magnetic triangular nanographenes}

\author{Nils Krane}
\email{nils.krane@empa.ch}
\author{Elia Turco}
\affiliation{nanotech@surfaces Laboratory, Empa - Swiss Federal Laboratories for Materials Science and Technology, 8600 Dübendorf, Switzerland}

\author{Annika Bernhardt}
\author{Michal Juríček}
\affiliation{Department of Chemistry, University of Zurich, Winterthurerstrasse 190, 8057 Zurich, Switzerland}

\author{Roman Fasel}
\affiliation{nanotech@surfaces Laboratory, Empa - Swiss Federal Laboratories for Materials Science and Technology, 8600 Dübendorf, Switzerland}
\affiliation{Department of Chemistry, Biochemistry and Pharmaceutical Sciences, University of Bern, 3012 Bern, Switzerland}

\author{Pascal Ruffieux}
\affiliation{nanotech@surfaces Laboratory, Empa - Swiss Federal Laboratories for Materials Science and Technology, 8600 Dübendorf, Switzerland}
\email{pascal.ruffieux@empa.ch}

\date{\today}

\begin{document}

\maketitle

\begin{abstract}
Inelastic electron tunneling spectroscopy (IETS) is a powerful measurement technique often used in scanning tunneling spectroscopy to probe excited states of various nanostructures, \textit{e.g.}, the magnetic properties of complex spin systems.
The observed excited states can be of magnetic and vibrational origin and it is therefore necessary to differentiate between these two excitation mechanisms.
Here, we investigate the spin $S=\nicefrac{1}{2}$ phenalenyl radical on Au(111).
IETS measurements feature inelastic excitations, whereas the spatial distribution of their intensity excludes any spin excitations.
Comparison to theoretical simulations proves the vibrational origin of those excitations and allows us to assign the observed features to distinct vibrational modes.

\end{abstract}

Open-shell nanographenes (NGs) have gained much interest in recent years as a versatile platform for entangled spin systems with long coherence times, due to the weak spin-orbit and hyperfine coupling of carbon-based materials~\cite{YazyevNanoLet2008,MinPRB2006,YazyevRPP2010,SlotaNature2018,OteyzaJPCM2022}.
Using a combination of in-solution and on-surface organic synthesis, such NG spin platforms can be built bottom-up from molecular precursors with various coupling motifs.
In this regard, the family of the zigzag-edged triangular NGs represent a prototypical group of carbon-based building blocks~\cite{PavlicekNatNano2017,MishraJACS2019,SuSciAdv2019,SuAngewandte2020,MishraNanoscale2021,SuNanoLett2021,TurcoJaAu2023}.
They are also commonly referred to as [$n$]triangulenes, with $n \geq 2$ being the number of benzene rings along the edge.
Due to the inherent sublattice imbalance of their structure, [$n$]triangulenes feature a total spin of $S=(n-1)/2$, scaling with the size of the triangulene~\cite{OvchinnikovTCA1978,LiebPRL1989}.
For these reasons, [$n$]triangulenes have been widely used as building blocks for various entangled carbon-based spin platforms, \textit{e.g.} as dimers~\cite{MishraAngewandte2020,KraneACSNano2018,SuNanoLett2020,TurcoArXiv2024}, trimers~\cite{ChengNatCom2022,DuNatCom2023}, cyclic structures~\cite{HieulleAngewandte2021}, one-dimensional chains~\cite{MishraNature2021, ZhaoArxiv2024} or two-dimensional networks~\cite{DelgadoArXiv2023,CatarinaPRR2023}.

Their complex magnetic ground and excited states are most often investigated by means of scanning tunneling microscopy (STM) and inelastic electron tunneling spectroscopy (IETS)~\cite{HeinrichScience2004}.
The underlying mechanism of the latter lies in the opening of a new tunneling channel by exciting the probed system inelastically, leading to steps in the differential conductance spectrum~\cite{JaklevicPRL1966,LambePR1968}.
However, the excited states do not necessarily have to be of magnetic origin as similar steps in the spectrum can also be caused by inelastic excitation of vibrational modes~\cite{StipeScience1998,HeinrichScience2002,HoJCP2002}.
When investigating magnetic properties of complex spin systems, it is therefore crucial to distinguish between these two excitation mechanisms and be aware for which vibrational modes significant IETS signals are to be expected.

\begin{figure}
    \centering
    \includegraphics[width=0.45\textwidth]{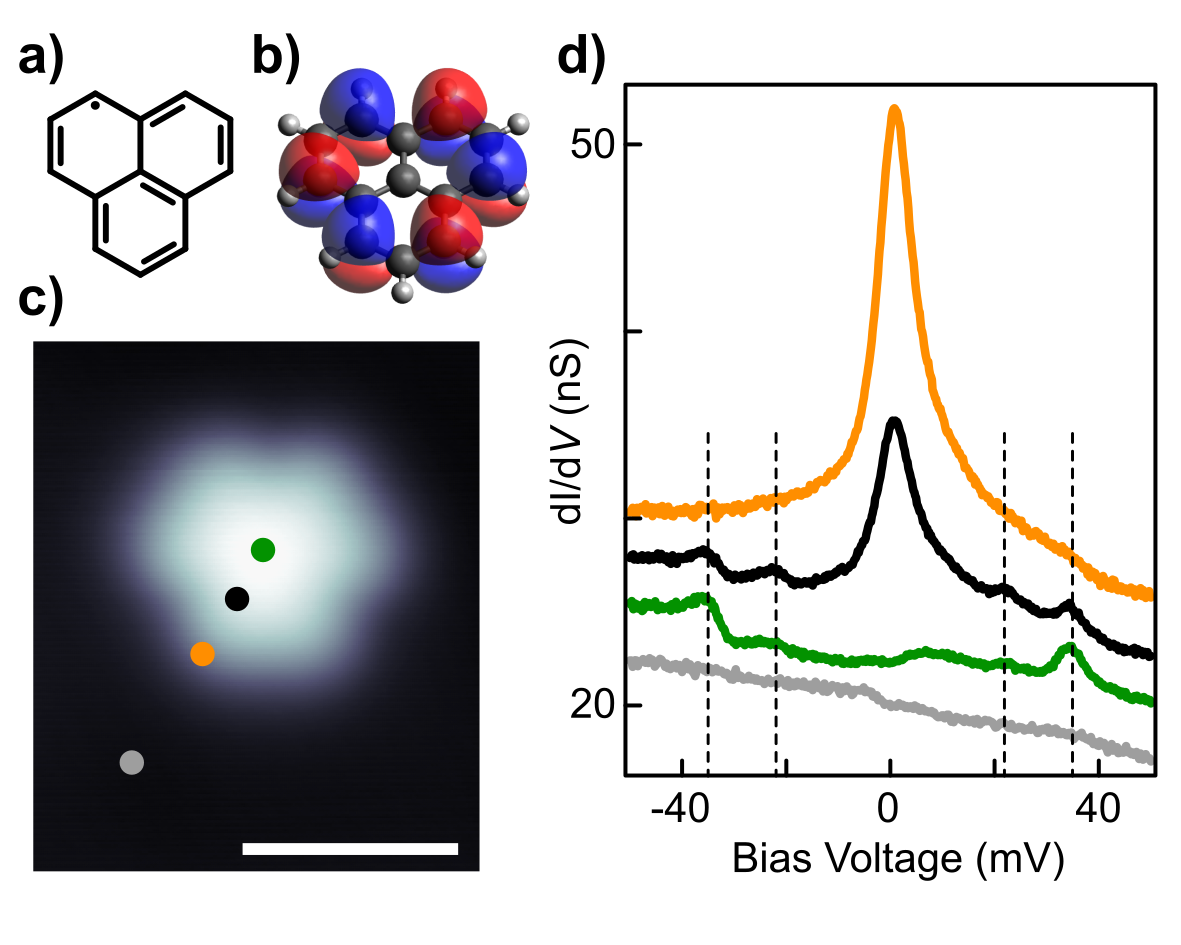}
    \caption{Chemical structure of the phenalenyl radical (a) and perspective view of its singly occupied molecular orbital (SOMO) plotted on an iso-density surface (b).
    (c) STM image of phenalenyl on Au(111). The scale bar indicates \SI{1}{\nano\meter}. Setpoint: \SI{-50}{\milli\volt}/\SI{1}{\nano\ampere}
    (d) \didv\ spectra taken at different positions on the phenalenyl, as well as, on the bare Au(111) as indicated in (c). Depending on the tip position some steps appear at different energies symmetric around zero bias voltage. The dashed vertical lines are drawn at \SI{\pm22.5}{\milli\volt} and \SI{\pm35}{\milli\volt}. Feedback opened at \SI{-50}{\milli\volt}/\SI{1}{\nano\ampere} and the spectra were offset in height for clarity.}
    \label{fig:Fig1}
\end{figure}

In this work, we investigate the smallest member of the [n]triangulene family, the phe\-na\-len\-yl radical (Figure~\ref{fig:Fig1}a), on a Au(111) surface, and demonstrate a simple method to identify vibrational excitations in IETS measurements.
By spatial mapping of the inelastic tunneling probability and comparison to simulations based on density functional theory (DFT) calculations, we are able to identify vibrational excitations and assign them to specific vibrational modes.

The synthesis of the hydrogen-passivated 1\textit{H}-phenalene is described in Ref.~\citenum{TurcoJaAu2023}. It was deposited under ultrahigh vacuum conditions from a quartz crucible held at \SI{70}{\celsius} onto an atomically clean Au(111) substrate held at room temperature and the sample subsequently transferred \textit{in-situ} into a low-temperature scanning tunneling microscope (LT-STM) with base temperature of \SI{4.5}{\kelvin}.
Bias voltages are applied to the sample and \didv\ spectra were obtained via lockin-amplifier with a voltage modulation of $V_\mathrm{mod} = \SI{0.5}{\milli\volt}$, if not stated otherwise.
The DFT calculations were performed for a single molecule in gas-phase using the Gaussian~16
package~\cite{Gaussian} with B3LYP functionals and 6-311g(d,p) basis-set.

On Au(111) the 1\textit{H}-phenalene can be deprotonated by tip-induced manipulation, in order to achieve the phenalenyl radical with spin $S=\nicefrac{1}{2}$ ground state~\cite{TurcoJaAu2023}.
The unpaired electron is located in the singly occupied molecular orbital (SOMO), which is distributed over the $p_z$-orbitals of the six outer carbon atoms, as depicted in Figure~\ref{fig:Fig1}b.
The interaction of the SOMO with the underlying Au(111) surface leads to an effective screening of the unpaired spin and the formation of a many-body singlet state, causing a characteristic Kondo resonance close to the Fermi energy~\cite{TurcoPRR2024}.
This Kondo resonance can be observed as a strong peak in the \didv\ spectrum taken at the edge of the phenalenyl (orange spectrum in  Figure~\ref{fig:Fig1}d), where the SOMO density is highest.

Figure~\ref{fig:Fig1}d depicts two \didv\ spectra taken at different locations on the phenalenyl radical, closer to the center of the molecule, showing a strong weakening of the Kondo peak.
This can be explained by a reduced coupling between tip and SOMO.
If the tip is placed right above the center of the molecule, all $p_z$ orbitals of the six outer carbon atoms, where the SOMO is located, are equally coupled to the tip. They are, however, interfering destructively due to opposite phases of the wave function and therefore quench the Kondo resonance completely (see green spectrum in Figure~\ref{fig:Fig1}d).

On the other hand, two new features can be observed when taking spectra close to the center of the phenalenyl. Two steps in the conductance appear symmetrically around zero bias at \SI{\pm35}{\milli\volt}. For the spectrum taken slightly off-center (black) there are also two additional steps at bias voltages of \SI{\pm22.5}{\milli\volt}.

The origin of these symmetric steps can be assigned to inelastic off-resonant tunneling, whereby the molecule is being excited, \textit{e.g.} vibrationally or magnetically, without changing its charge state.
At a certain threshold bias voltage, which corresponds to the energy of the excited state, a new channel is opened for electrons to tunnel inelastically through the molecule to the substrate~\cite{JaklevicPRL1966,LambePR1968}. This new channel appears as two steps in the \didv\ spectrum at energies symmetrically around zero bias voltage.

For a single phenalenyl radical on Au(111), a spin excitation can be ruled out as the origin, since a single unpaired spin-$\nicefrac{1}{2}$ without applied magnetic field is not expected to have an excited spin state.
In addition, spin excitation requires tunneling through the molecular orbital carrying the spin (the SOMO).
If the inelastic steps were spin excitations, they should only occur at tip positions where tunneling into the SOMO is possible, in this case indicated by the intensity of the Kondo resonance.

\begin{figure}
    \centering
    \includegraphics[width=0.45\textwidth]{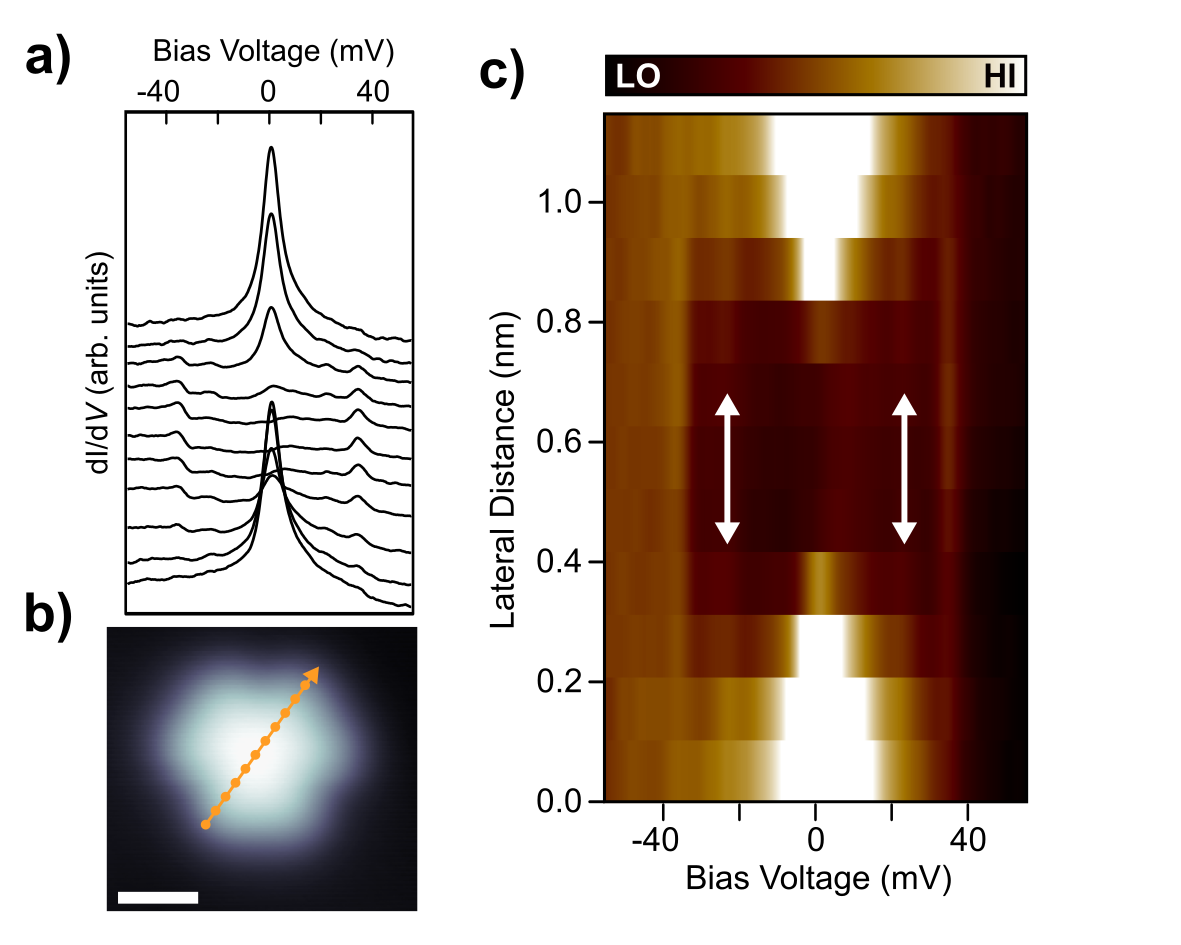}
    \caption{
    \didv\ spectra taken along a line across a phenalenyl molecule. (a) The Kondo resonance is strongest when the tip is positioned at the edges of the phenalenyl, whereas the inelastic steps appear in the center. Feedback opened for each spectrum at \SI{-60}{\milli\volt}/\SI{1}{\nano\ampere}. The spectra are vertically offset for clarity.
    (b) STM image of phenalenyl, indicating the positions where spectra were acquired. The scale bar indicates \SI{0.5}{\nano\meter}. Setpoint: \SI{-60}{\milli\volt}/\SI{1}{\nano\ampere}
    (c) Same spectra as in (a) but normalized, such that conductance at set point \SI{-60}{\milli\volt} is set to one for all spectra. The color scale is clipped for the Kondo resonance for better highlighting of the inelastic steps.
    }
    \label{fig:Fig2}
\end{figure}

The intensities of the observed inelastic steps in the \didv\ spectra, however, are following a different spatial distribution than the Kondo peak, as is depicted in Figure~\ref{fig:Fig2} for a line of 11 \didv\ spectra, taken across the phenalenyl molecule.
As expected, the Kondo resonance is most prominent at the edges of the molecule, where the tip is placed right above one of the $p_z$-orbitals of the SOMO, minimizing the destructive interference with the other $p_z$-orbitals.
In contrast, the steps at \SI{\pm35}{\milli\volt} are strongest in the center of the molecule and fade to the outside, whereas the steps at \SI{\pm22.5}{\milli\volt} (indicated by the white arrows in Fig.~\ref{fig:Fig2}c) appear in an intermediate area but vanish towards the center.

The observed spatial distribution of the inelastic steps hints towards vibrational excitation as the underlying mechanism.
The mixing of Kondo resonance and vibrational excitation leads to overshooting at the conductance steps~\cite{ParksPRL2007,FernandezPRL2008,EickhoffPRB2020,KorytarPRB2021,SuNanoLett2021}, similar in appearance to higher-order spin-scattering observed for magnetic excitations~\cite{TernesNJP2015}. 

In the case of  off-resonant vibrational excitation, the tunneling probability is not proportional to the overlap between the tip and unperturbed wavefunction of the molecular orbital $\Psi_m$.
Instead, it is proportional to the overlap between tip and a vibrationally perturbed wavefunction~\cite{LorentePRL2000} $\delta\Psi_{m,k} = \delta Q_k (\partial \Psi_m / \partial Q_k)$, where $\delta Q_k = \sqrt{\hbar/2\omega_k}$ is the root mean square displacement of the mass weighted normal coordinate $Q_k$ of vibrational mode $k$. 
Consequently, the spatial distribution of the IETS steps' intensity differs to that of the elastic tunneling into the corresponding orbital.
As a rule of thumb and assuming an $s$-wave tip, a vibrational mode has to be of similar (local) symmetry as the phase of the wave function in order to observe a significant off-resonant vibrational assisted tunneling intensity~\cite{PavlicekPRL2013,ReechtPRL2020}.

\begin{figure}
    \centering
    \includegraphics[width=0.45\textwidth]{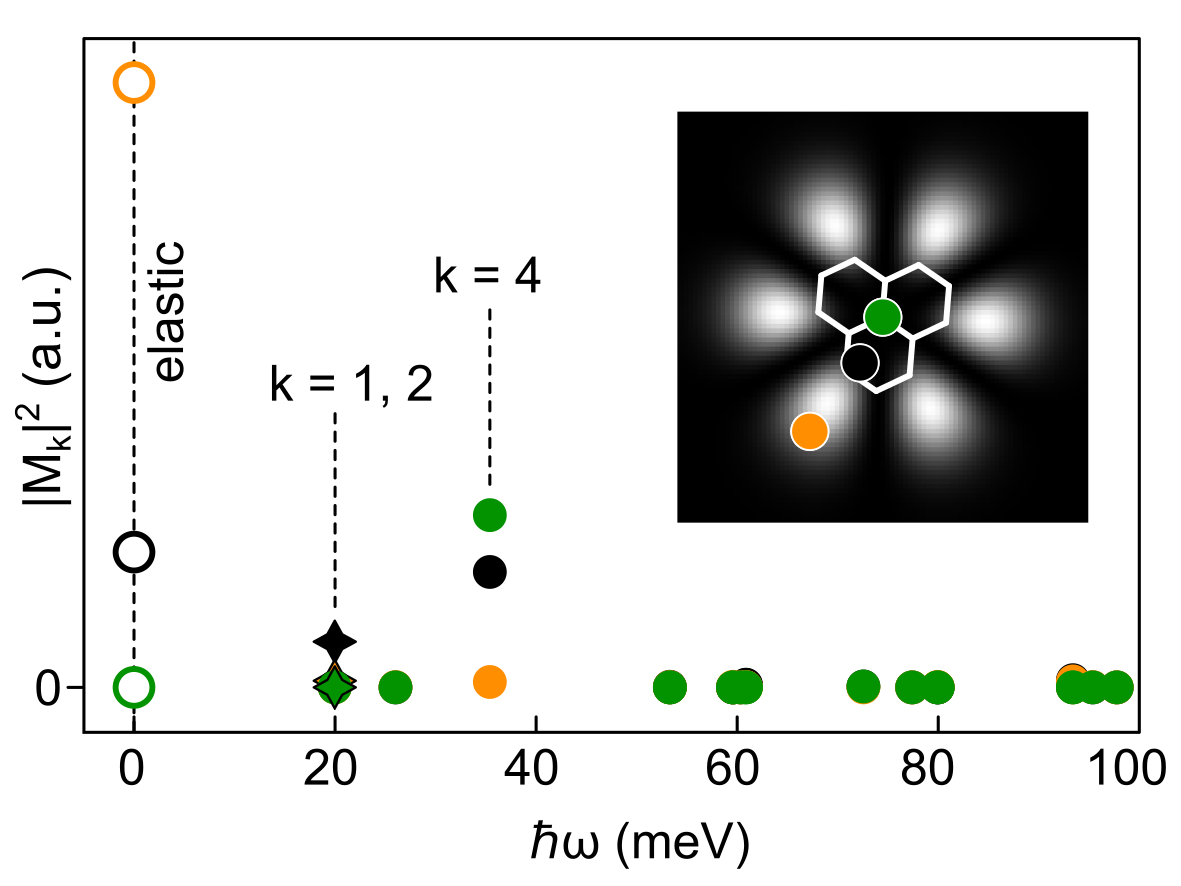}
    \caption{Tunneling matrix elements $\left | M_k \right |^2$ for all vibrational modes $\hbar\omega < \SI{100}{\milli\electronvolt}$ (full circles), as well as elastic tunneling (hollow circles), for three different tip positions (see inset). Only the modes $k=1, 2$ (\SI{20}{\milli\electronvolt}) and $k=4$ (\SI{35}{\milli\electronvolt}) show significant intensities.
    The markers for mode $k=1$ are displayed as stars for better distinction from $k=2$.
    Inset: Simulated \didv\ map of elastic tunneling matrix elements into the SOMO. The colored circles indicate tip position of the calculated vibrational excitation intensities in the main graph.
    }
    \label{fig:Fig3}
\end{figure}

The electron-phonon coupling for the off-resonant vibrational excitation probability can  be simulated using DFT calculations, \textit{e.g.}, following the description from Ref.~\cite{ReechtPRL2020}. For each vibrational mode $k$, the structure of phenalenyl is perturbed by $\pm\delta Q_k/2$ and the corresponding change in wavefunction calculated by  $\delta\Psi_{m,k} \approx \Psi_m \left (+\delta Q_k/2 \right ) - \Psi_m \left (-\delta Q_k/2 \right )$.
The tunneling probability is then obtained using Bardeen's theory of tunneling~\cite{Bardeen1961}:
\begin{equation}
\left | M_k \right |^2= 
\left | \frac{\hbar^2}{2m_e}\int \mathrm{d}\vec{S}\left [ \Psi_s \vec\nabla \delta\Psi_{m,k} - \delta\Psi_{m,k} \vec\nabla \Psi_s \right] \right |^2,    
\end{equation}
assuming an \textit{s}-wave orbital $\Psi_s$ for the tip with workfunction $\phi=\SI{5}{\electronvolt}$ at a height $z_t=\SI{7}{\angstrom}$ above the molecule. The integrataion plane $\vec{S}$ was chosen to be at $z_S=\SI{1.6}{\angstrom}$.
The obtained tunneling matrix elements $\left | M_k \right |^2$
for the low energy vibrational modes ($\hbar\omega < \SI{100}{\milli\electronvolt}$) at three different tip positions are displayed in Figure~\ref{fig:Fig3}.

\begin{figure*}
    \centering
    \includegraphics[width=0.9\textwidth]{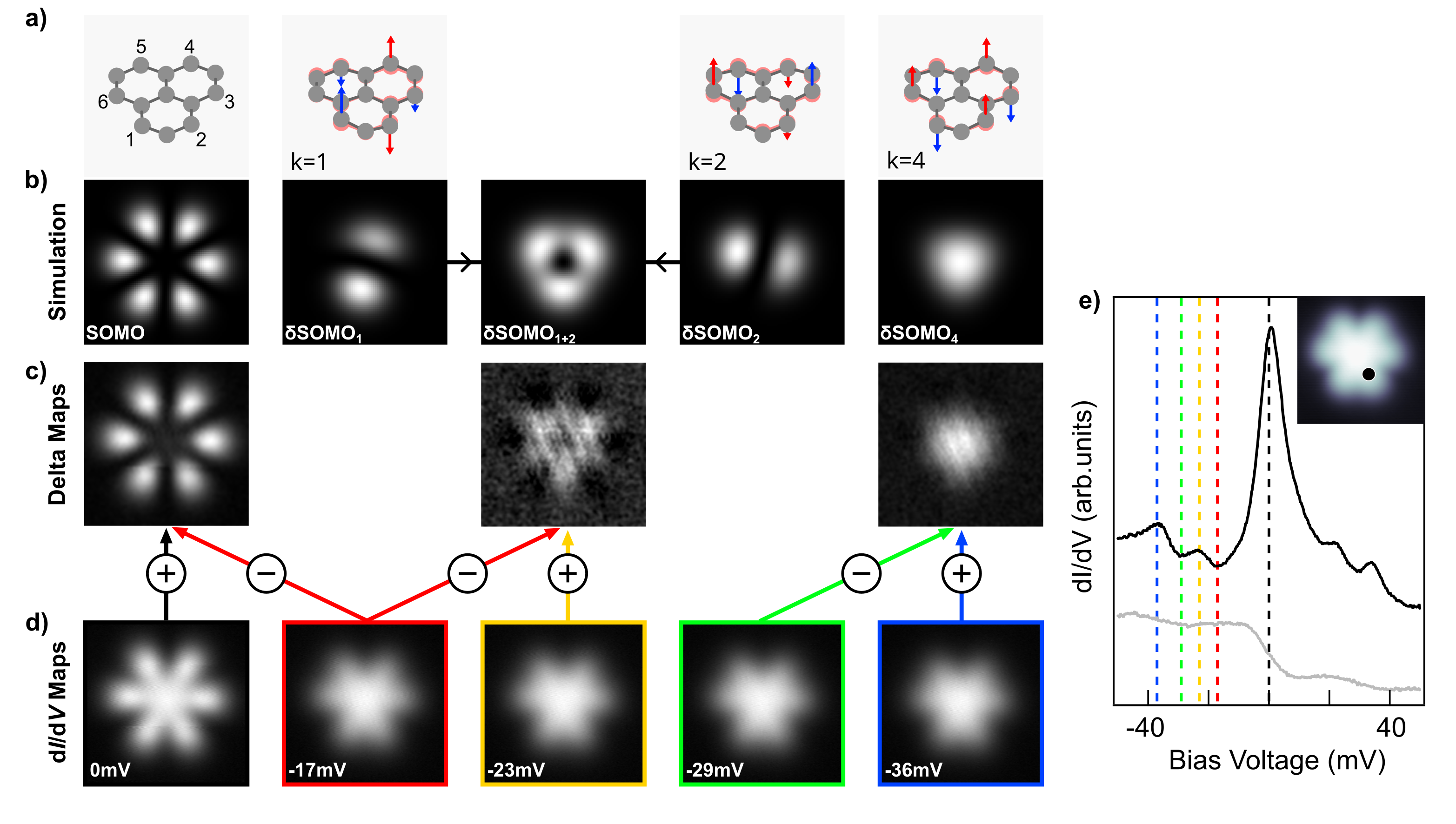}
    \caption{Spatial mapping of vibrational excitation intensities. 
    (a) Structure of phenalenyl in equilibrium position (left) as well as dislocated structures corresponding to vibrational modes $k=$1, 2 and 4. The arrows indicate the direction of movement for certain atoms. The red and blue color of the arrows indicate the phase of the SOMO at the corresponding atom. The displacement of the atoms is exaggerated for better visibility.
    (b) Simulated \didv\ maps for elastic (left) and inelastic tunneling (center and right). The maps for the two degenerate modes $\delta\mathrm{SOMO}_1$ and $\delta\mathrm{SOMO}_2$ are displayed individually, as well as, the sum of their intensities $\delta\mathrm{SOMO}_{1+2}$.
    (c) Experimental maps displaying spatial distribution of elastic tunneling and vibrational excitation intensity, obtained by subtraction of constant height maps in (e).
    (d) Constant height \didv\ maps taken at various bias voltages, as indicated by dashed lines in (e). Feedback was opened at the center of molecule (\SI{-50}{\milli\volt}/\SI{0.5}{\nano\ampere}) and $V_\mathrm{mod} = \SI{2.5}{\milli\volt}$. Size of maps: 1.5\,x\,\SI{1.5}{\nano\meter\squared}.
    (e) \didv\ point spectrum taken with same parameters as the maps in (d) at tip position marked in the inset.
    }
    \label{fig:Fig4}
\end{figure*}

A non-vanishing vibrational exctitation intensity is found for three vibrational modes, which are also sketched in Fig.~\ref{fig:Fig4}a.
The first two modes ($k=1,2$) are degenerate in energy ($E_{1,2}=\SI{20.0}{\milli\electronvolt}$) and have significant contribution in the intermediate area between center of molecule and its outer edge.
Vibrational mode $k=4$, on the other hand, shows a pronounced intensity localized in the center of the molecule with an energy of $E_4 = \SI{35.4}{\milli\electronvolt}$.

All three modes with significant electron-phonon coupling comprise of an out-of-plane motion including the six outer carbon atoms where the SOMO is localized.
In the case of the $k=4$ mode, the displacement of the carbon atoms is of the same symmetry as the phase of the SOMO, as shown in the rightmost panel of Figure~\ref{fig:Fig4}a, whereby the color of the arrows indicates the phase of the SOMO wavefunction.
The motion of the vibrational mode breaks the symmetry of the SOMO and therefore reduces the deconstructive interference of the six $p_z$ orbitals, leading to a high tunneling probability for a tip placed in the center of the molecule.
Similar observations can be made for the two degenerate modes $k=1$ and 2,
where the dislocations of the carbon atoms match the phase symmetry on a local level.
For example, mode $k=1$ features two different local regions in which the dislocation follows the symmetry of the orbital phase: The first one comprises the two carbon atoms labeled 1 and 2 with blue pointing upward and red pointing downwards. The second region includes atoms 3, 4 and 5 and the direction of the colored arrows is inverted with red pointing upwards and blue pointing down.
This inverted direction of arrows with same color indicate, that the two regions are antisymmetric with respect to each other, leading to a vanishing inelastic tunneling probability when a tip is placed between these two regions.
Analogous observations can be made also for mode $k=2$.

Figure~\ref{fig:Fig4}b shows the simulated spatial distribution of the inelastic excitation intensities for the three significant vibrational modes, as well as elastic tunneling into the SOMO.
The individual maps $\delta\mathrm{SOMO}_{1}$ and $\delta\mathrm{SOMO}_{2}$ of the two degenerate modes $k=1,2$ are not threefold symmetric, but feature a nodal plane between the two local symmetry regions.
However, the sum of their intensities $\delta\mathrm{SOMO}_{1+2}$ has again a threefold symmetry and appears as slightly triangular ring with vanishing intensity in the center.
For the $k=4$ mode, the inelastic tunneling is as expected strongest in the center of the molecule.
One can clearly see that neither mode is as delocalized as the elastic tunneling into the SOMO. Therefore, no significant electron-phonon coupling is expected when the tip is positioned at the edges of the phenalenyl. This allows to take spectra without any significant inelastic vibrational excitations.

In order to fully confirm the vibrational origin of the observed steps in the \didv\ spectra, we experimentally mapped their spatial distribution.
For this, we recorded constant height \didv\ maps at bias voltages just above and below the energy threshold for each inelastic feature (Figure~\ref{fig:Fig4}d,e).
After a lateral drift compensation and numerical smoothing, the lower energy maps (taken "before" the step) were subtracted from the higher energy maps (taken "above" the step) to obtain a spatial mapping of the corresponding step intensity.
We also recorded one map at zero bias voltage and subtracted the map taken at \SI{-17}{\milli\volt} in order to obtain a mapping of the elastic Kondo resonance for comparison.
The resulting three maps are displayed in Figure~\ref{fig:Fig4}c and are in very good agreement with the simulated spatial distributions.

Please note that the Kondo resonance is causing a non-constant background for small bias voltages, in particular for the lower energy step at \SI{\pm22.5}{\milli\volt}.
The subtraction of the two maps taken at \SI{-23}{\milli\volt} and \SI{-17}{\milli\volt} therefore overestimates the Kondo background and causes the six darker areas of reduced conductance around the bright triangular ring (Figure~\ref{fig:Fig4}c middle).
In contrast, the area of reduced conductance in the center of the ring cannot be ascribed to this background overestimation, since the Kondo resonance does not have any intensity there, and represents the actual spatial distribution of the inelastic tunneling probability.
The conformity between simulated and measured maps accordingly proofs that the observed steps in the \didv\ spectra are indeed caused by off-resonant vibrational excitations.

In conclusion, we successfully demonstrated inelastic vibrational tunneling through a phenalenyl radical, by experimentally mapping the spatial distributions of steps in the differential conductance.
Comparison with \textit{ab-initio} DFT calculations allowed us to ascribe the inelastic features to the excitation of three out-of plane vibrational modes.
The identification of off-resonant vibrational excitation is an important responsibility to distinguish vibrational IETS from other inelastic tunneling mechanisms, \textit{i.e.} spin excitations.
Understanding the spatial distribution of the electron-phonon coupling is also crucial for selecting suitable tip positions when investigating magnetic properties of nanographenes with IETS.

% ##########################################################################

\section{Acknowledgments}
This research was supported by the Swiss National Science Foundation (grant no. 200020\_175923 and CRSII5\_205987) as well as by the ITN Ultimate Program (813036).
We also greatly appreciate financial support from the Werner Siemens Foundation (CarboQuant).
Skillful technical assistance by Lukas Rotach is gratefully acknowledged.
For the purpose of Open Access, the author has applied a CC BY public copyright license to any Author Accepted Manuscript version arising from this submission.

% ##########################################################################

\bibliography{references}

\end{document}